\documentclass[conference]{IEEEtran}
\IEEEoverridecommandlockouts
\usepackage{cite}
\usepackage{amsmath,amssymb,amsfonts}
\usepackage{algorithmic}
\usepackage{graphicx}
\usepackage{textcomp}
\usepackage{xcolor}
\usepackage{tabularx,graphicx}
\usepackage[T1]{fontenc}
\usepackage{longtable, tabu}
\usepackage{float}
\usepackage{placeins}
\def\BibTeX{{\rm B\kern-.05em{\sc i\kern-.025em b}\kern-.08em
    T\kern-.1667em\lower.7ex\hbox{E}\kern-.125emX}}

\usepackage{array}% in the preamble
\newcolumntype{C}[1]{>{\centering\let\newline\\\arraybackslash\hspace{0pt}}m{#1}}
\usepackage[utf8]{inputenc}

\usepackage[colorinlistoftodos,prependcaption]{todonotes}
\usepackage{blindtext}
\presetkeys%
    {todonotes}%
    {inline,backgroundcolor=yellow}{}

\begin{document}

\title{Multi-Approach Debugging of Industrial IoT Workflows}

\author{\IEEEauthorblockN{Andreia Rodrigues}
\IEEEauthorblockA{\textit{DEI} \\
\textit{Faculty of Engineering}\\ \textit{University of Porto}\\
Porto, Portugal\\
up201404691@fe.up.pt}
\and
\IEEEauthorblockN{Jos\'{e} Pedro Silva}
\IEEEauthorblockA{\textit{Critical Manufacturing} \\
Porto, Portugal\\
josesilva@criticalmanufacturing.com}
\and
\IEEEauthorblockN{Jo\~ao Pedro Dias}
\IEEEauthorblockA{\textit{INESC TEC and }\textit{DEI} \\
\textit{Faculty of Engineering}\\ \textit{University of Porto}\\
Porto, Portugal\\
jpmdias@fe.up.pt}
\and
\IEEEauthorblockN{Hugo Sereno Ferreira}
\IEEEauthorblockA{\textit{INESC TEC and }\textit{DEI} \\
\textit{Faculty of Engineering}\\ \textit{University of Porto}\\
Porto, Portugal\\
hugosf@fe.up.pt}
}

\maketitle

\begin{abstract}
Industrial Internet-of-Things (IIoT) results from the addition of sensing and actuating capabilities to industrial environments to improve the overall manufacturing processes. Some of these systems have highly-complex tasks of monitorization and control and need to be programmed accordingly. The use of visual programming, such as workflows, is common in these systems due to the abstraction they provide to the systems programmer. However, such programming environments have several deficiencies on what regards debugging capabilities, mostly due to the constraints that difficult the use of traditional mechanisms. The work presented in this paper approaches these issues, delving into the design and implementation of a multi-strategy debugging mechanism into a commercial-grade Manufacturing Execution System. To validate the approach, a proof-of-concept was then developed and validated against different debugging scenarios.
\end{abstract}


\section{Introduction}

Ever since the Industrial Revolution, increasing efficiency in manufacturing systems has been a constant endeavour. Companies thrive to improve productivity and reduce costs without compromising quality, enhancing the efficiency of the manufacturing process.

The birth of Internet-of-Things followed by its application to industrial environments, \textit{viz.} Industrial Internet of Things (IIoT), pushes the efforts of improving the manufacturing processes even further, towards the so-called Industry 4.0. The use of networked sensors and intelligent devices to collect data about the manufacturing processes, boost operational efficiency and reduce detection and troubleshooting time, hence resulting in overall money and time savings~\cite{54}. The computational elements that control physical entities, using sensors and actuators feedback in real-time, enable the detection and prevention of failures to avoid significant losses~\cite{55}. The software controlling these devices requires support for management, configuration, control and debug of operations, especially during the maintenance process.

% \todo{Somewhere tens de introduzir o que é um workflow, que tipo de programação é, que se baseia em metaforas visuais, visto que é algo \textit{core} aos teu trabalho. Relacionar com PLC programming p.ex., mesmo que tenhas isso explicado mais a frente, aqui tem de aparecer uma quick notion.}

Maintenance of these systems is highly required to ensure that the software product delivered to the user still maintains a high level of quality and satisfies the client's requirements. It is greatly assisted with suitable debugging techniques that ensure the software's sustainability and quality, helping to identify the causing defects related to software failures.

The aforementioned debugging needs of IIoT systems still lack in terms of tools, approaches and methodologies that can be used for such~\cite{8411738}. The problem grows when considering that, commonly, these systems are programmed using \textit{visual workflows}, that typically lack tools for debugging (as further explored in the next Sections). These workflows are an abstraction of the different tasks to be executed by the equipment, creating blocks that aggregate some of the steps to be performed, receiving inputs from the previous tasks and delivering outputs to be used by the following.

The main goal of this work is, by studying the best practices of debugging different kinds of systems, to design and develop a debugging approach (and protocol) that can be used for IIoT and meet the requirements of a real-world Manufacturing Execution System.  A proof-of-concept was implemented on top of the Critical Manufacturing Execution System and their IoTMT solution (\emph{c.f.} Section~\ref{ssec:4b}), allowing us to run a set of preliminary tests to verify the viability of the approach and the underlying protocol.

The paper is structured as follows: Section~\ref{sec:background} presents contextualization of the main concepts related with this work, Section~\ref{sec:relatedwork} presents the related available work, Section~\ref{sec:problemStatement} presents the solution proposed by this article, Section~\ref{sec:proposedSolution} details the proposed solution, Section~\ref{sec:experiments} lists the simulated scenarios used to validate our approach and, finally, Section~\ref{sec:conclusion} presents the main contributions and conclusions of this work.

\section{Background}
\label{sec:background}

%-----------------------------------------------------------------------------

This section provides an overview
of the key concepts regarding the approach presented in this article.

\subsection{Industry 4.0} %------------------------------------------------------------------------

Industry 4.0 is an emerging concept that aims at a higher level of operational efficiency and productivity in factories by augmenting the level of automation in manufacturing systems~\cite{1}. It is defined as \emph{``the integration of complex physical machinery and devices with networked sensors and software, used to predict, control and plan for better business and societal outcomes''}~\cite{2}. It unifies the concept of objects, machines, assembly lines and whole factories~\cite{3} leading to the ``smart factory'' vision with the implementation of cyber-physical systems (which \emph{``extend real-world physical objects by interconnecting them all together and providing their digital descriptions''}~\cite{8}). Factories become more intelligent, flexible, dynamic and autonomous, leading to the growth of productivity, reduction of costs, increase of effectiveness and efficiency and improvement of the overall quality of the products~\cite{1}.

The Internet-of-Things is an emerging paradigm where everyday objects can be equipped with identifying, sensing, networking and processing capabilities that will allow them to communicate with one another and with other devices and services over the Internet to accomplish some objective~\cite{4}. It conceptualizes a world where all things are wirelessly and seamlessly connected and can be controlled remotely and exchange data at any time~\cite{5}.

IoT devices are equipped with embedded sensors, actuators, processors, and transceivers~\cite{2,6}. Sensors and actuators are devices that help to interact with the physical environment. While sensors provide inputs about the device's current state (internal state and environment), an actuator performs actions to affect the environment or device in some way~\cite{4,6}. The combination of these elements can enable objects to simultaneously be aware of their environment and interact with people, both goals of IoT.

The subset of IoT specific to industrial applications (Industrial Internet-of-Things) is where IoT and Industry 4.0 meet. Smart and efficient manufacturing can be achieved with IIoT, connecting the shop-floor to production management~\cite{7}.

It focuses on the manufacturing stage of the product's life cycle, aiming for a quick and dynamic response to demand changes~\cite{8} by equipping high-tech products such as sensors and actuators, software and wireless connectivity in the production equipment~\cite{1}. These generate large amounts of data that can grant a better understanding of the manufacturing process and enable a more efficient and sustainable production system, reducing costs without compromising quality~\cite{8,9}. An example of this is the machinery maintenance process, as analyzing machine data will allow to understand and identify the top causes of failure and predict component failures to avoid unscheduled machine downtime, which usually leads to company resource losses~\cite{9}.
 
 \subsection{Manufacturing Execution Systems}

 %----------------------------------------------------

A Manufacturing Execution System (MES) is \emph{``an information system that drives the execution of manufacturing operations''}~\cite{11}. It focuses on achieving and maintaining high performance of production~\cite{12} through the digitalization of shop-floor activities with the collection, analysis, and exchange of information captured in real-time during this process~\cite{10}.

According to MESA\footnote{MESA stands for Manufacturing Enterprise Solutions Association}, a manufacturing execution system is a dynamic information system that ensures the effective execution of the manufacturing operations through the gathering of real-time data, guiding, triggering and reporting on shop-floor activities as events occur. It \emph{``manages production operations from the point of order release into manufacturing to the point of product delivery into finished goods''}~\cite{13}.

\begin{figure}[htbp]
\centerline{\includegraphics[width=0.5\textwidth]{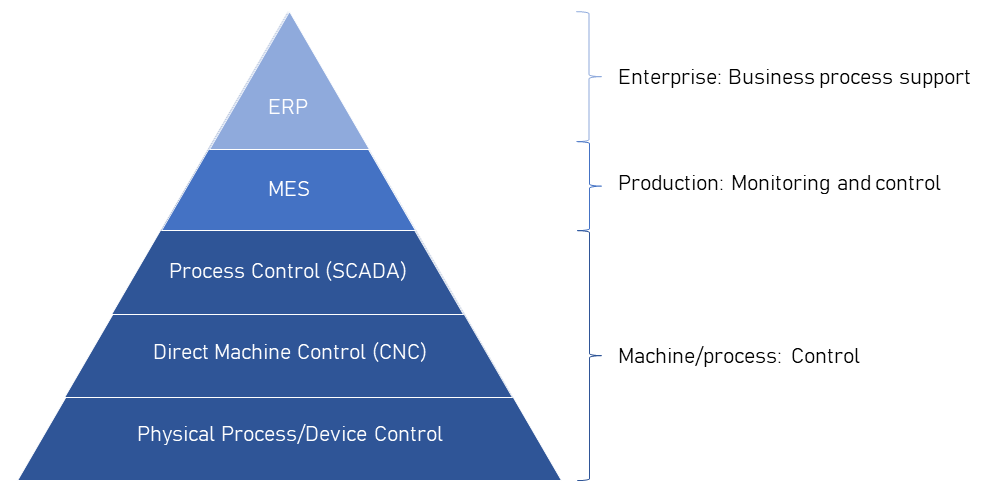}}
\caption{Automation pyramid (Adapted from:~\cite{10}).}
\label{fig:pyramid}
\end{figure}

This type of system focus on the vertical integration of manufacturing processes with business processes~\cite{10}, working as a middle-layer, by bridging enterprise information systems (ERPs) with the shop-floor equipment, implementing manufacturing operational management (MOM) functions in the enterprise~\cite{14}. This is illustrated in Fig.~\ref{fig:pyramid}.

At Critical Manufacturing, this system is responsible for performing the following set of tasks~\cite{12}:
\begin{itemize}
\item Monitoring and assuring the correct execution of the production process;
\item Monitoring and controlling the material used in the production process;
\item gathering of information about the production process;
\item Providing the tools for the analysis of the data obtained, related to the production process, to optimize efficiency;
\item Delivering and managing work instructions;
\item Providing the tools to solve problems that may come up during production and optimize procedures.
\end{itemize}

The analysis of the data collected throughout the product life cycle is of growing importance for organizations. MES is responsible for this analysis, followed by the integration and presentation of the results in the industrial production, providing real-time, accurate, granular data which allows employees to have a better insight into the manufacturing processes, optimizing their decisions for controlling and quickly react to any issues that may arise during production~\cite{10,15}.

Manufacturing Execution Systems play a critical role in Industry 4.0, as it \emph{``accommodates the Industrial Internet-of-Things (IIoT)-enabled production marketplace''}, playing a vital role as an enabler of further innovation in manufacturing~\cite{12}.
 
\subsection{Software Development Life Cycle in Industrial Systems} %----------------------------

The increasing usage of physical entities equipped with sensors and actuators in the manufacturing industry, through the integration of IoT in factories, provides meaningful feedback in real-time across a virtual network regarding the manufacturing process, enabling the fast detection and prevention of machinery failures~\cite{1}. Known as  Programmable Logic Controllers (PLC), these computational entities are widely used in automation control and will autonomously perform many processes within cyber-physical systems~\cite{20}.

PLCs are \emph{``computer-based, solid-state, single processor devices''} capable of controlling many types of industrial equipment and entire automated systems~\cite{20}. They are characterized by their cyclic data processing behaviour which consists of reading all input values (provided by sensors), executing the PLC program with the received values and, when finalized, writing all output variables (which control the actuators), later restarting the cycle~\cite{21}. 

These controllers are cheap, highly efficient and reliable in applications which involve sequential control and synchronization of processes~\cite{20}, hence the reason why they are the standard industrial platform nowadays~\cite{21}. However, as the complexity of these controllers increases, the more difficult it is to ensure their reliability~\cite{22}.

PLCs have been, historically, programmed using visual notations, with schematic or ladder diagrams instead of usual computer languages~\cite{20}. Examples of these graphical languages are Ladder Diagrams (LD), Function Block Diagrams (FBD), and Sequential Function Charts (SFC)~\cite{21}.

%\begin{figure}[htbp]
%\centerline{\includegraphics[width=0.4\textwidth]{images/V.png}}
%\caption{SDLC V-Model (Source: Wikipedia).}
%\label{fig:v-model}
%\end{figure}

The SDLC model most commonly used for developing PLC-based control systems is the V-model, where validation, verification, and testing are planned and usually performed in parallel with the requirement gathering, design, and implementation phases, respectively. It has the main advantage of including some validation before the development phase has begun. However, if errors are found during the testing phase, the test documents, along with requirement documents, will have to be updated, which can be quite expensive with the increase of the project complexity~\cite{22}.

\subsubsection{Software Maintenance} %-------------------------------------------------------------

Software maintainability is \emph{``the ease with which a software system can be modified to correct faults, improve performance or other attributes or adapt to a change of environment''}~\cite{23}. It takes place once the software product has been delivered to the user~\cite{24} and can involve repair/modification of the software (bug fixes), implementation of new requirements, or adaptive maintenance of the environment where the software is operating~\cite{21}. %(Fig.~\ref{fig:maintenance})

% \begin{figure}[htbp]
% \centerline{\includegraphics[width=0.4\textwidth]{images/maintenance.png}}
% \caption{Example of the maintenance process. (Source: ~\cite{24}).}
% \label{fig:maintenance}
% \end{figure}

Ensuring software quality has become a central issue because most organizations rely on software products to run their business operations efficiently and effectively to remain competitive. Thus it is crucial to ensure the sustainable quality of a software product throughout its life cycle.  To ensure this a good maintenance process is required~\cite{24}.

The maintenance phase ensures that the delivered software product sustains a high level of quality and satisfies the client's requirements~\cite{24}, aiming at adapting or perfecting the system towards this goal~\cite{25}. This process is invoked both when there is a change in the software requirements, and new features need to be implemented, or when failures are detected, and the software needs to stay operational~\cite{26}. Maintainability is one of the software's quality factors, as a good maintenance process ensures a successful service in the long run~\cite{24}.

\subsubsection{Software Maintenance in Industrial Systems} %---------------------------------------

Factories require consistent professional maintenance of automated production systems. The complexity of these systems, both hardware and software, has been increasing, and they often require a life-span of +10 years. For this reason, maintaining the developed software is strongly required to ensure the software's quality and reliability~\cite{21}.

%The software used in these production systems has special requirements on the development and maintenance process because it is strongly influenced by the integration of sensors and actuators in the hardware~\cite{28}. The maintenance phase takes advantage of the implementation of cyber-physical systems (CPS), using it as a service support system. The physical devices capture information in real-time that will allow to monitor a machine's health condition, know if the solution is operating according to the client's requirements, and detect failures as they happen. It will also allow detecting failures that may potentially happen in the future, using preventive maintenance. For this reason, these devices are considered to be at the core of new industrial machinery maintenance due to their ability to enable life cycle observation~\cite{21,28}.

Special requirements on the development and maintenance process are also needed, due to the need for rapid adjustment of production capacity and functionality, in response to new market conditions. With a fast-changing market and customers demanding a higher level of product customization, industrial software must be updated continuously and re-configured to cope with changes to production processes or the introduction of new products. This imposes pressure over the need for software development and maintainability, demanding faster development processes while maintaining software quality~\cite{56}.

\subsection{Debugging} %---------------------------------------------------------------------------

Software debugging is the process of identifying errors or defects in software or a computer system and solving them so that the program works correctly accordingly to specification~\cite{31}. Any step of a program that performs unexpectedly is termed to be a fault. Debugging is an essential part of the software engineering process, being an arduous, time-consuming, costly task~\cite{30}.

Human-made software should not be considered to be reliable, safe, secure, or always available unless it is thoroughly tested and verified, due to the unavoidable presence or occurrence of faults~\cite{25}. Testing these systems is an important part of the software development life cycle. After, or even during, the development phase, testing and debugging should be taken seriously and given high priority. Without these steps, high-quality, reliable software cannot be provided in today's fast track-based world~\cite{30}. The severity of a defect depends on the failure domain, controllability, and consistency of the failures encountered and the consequences in the environment. The time required to solve a bug is directly related to the complexity of the error~\cite{25}.

%As the total number of faults encountered increases, so does the cost of software development. Something that helps to decrease this cost is to detect and locate faults in the software in the early stages of development~\cite{30}. This effort reduces the costs of maintenance and accelerates the development of the software (as less time will be wasted on testing).

A debugger is a tool that allows debugging a program, to see what is going on ``inside'' a program while it executes or what it was doing at the moment it crashed. Any application software will inevitably contain bugs during the development cycle. The distance from the defect causes to the failure may be substantial in both time and space, so developers require access to powerful debugging tools for the correction of these software flaws, allowing them to work more efficiently and to better dig into the detailed operation of their application~\cite{32,33}.

\subsubsection{Remote Debugging} %-----------------------------------------------------------------

Traditionally, software applications are debugged in an environment in which the debugger is executing on the same computing device as the application being debugged. In some cases, resources such as memory, processing power, and network are consumed by the installation and execution of development and debugging environment (e.g. via an IDE). This is a critical factor because such computing devices may have limited resources, e.g., storage, processing and communication~\cite{34}. In these cases, using traditional debugging techniques may not be an option, and a remote debugger seems to be a reasonable solution.

\begin{figure}[htbp]
\centerline{\includegraphics[width=0.4\textwidth]{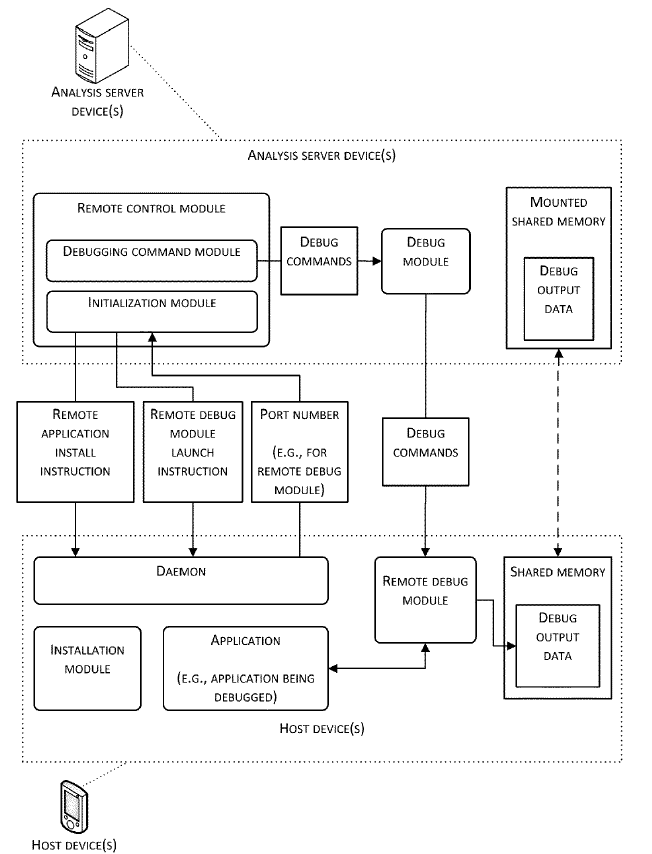}}
\caption{Example of a Remote Debugging approach~\cite{34}.}
\label{fig:remotedebug}
\end{figure}

In the remote debugging, or cross-debugging, technique, the debugger runs on a host machine, while the program to be debugged is running on a specific hardware platform of a target machine (Fig.~\ref{fig:remotedebug}). The application is launched using a remote debug module executing on the host machine, and it communicates with the target machine via a communication channel that can be set through a serial port, parallel port or network card interface, taking control of this machine to run the program~\cite{33, 34}.

During a debugging session, debug commands are sent from the host machine to the target program and these commands are employed for debugging the application on the remote host device. The commands may include the launch of the application to be debugged, setting breakpoints for pausing the execution of the application at a certain execution point, resuming execution following a breakpoint, stepping through instructions of the application or terminating it. Debug commands can also \textit{ask} for information regarding the state of the application such as the values stored in variables, parameters, registers and program counters of the application, stack traces for the call stack of the application, dumps of active memory following a crash or failure and such~\cite{34}. This debugging technique is especially useful for debugging embedded software, and it requires cooperation between the host and target machines~\cite{35}.

\subsubsection{Debugging on Industrial Systems} %--------------------------------------------------

Companies are competing fiercely to provide high-quality software at the lowest possible cost. Software maintenance plays a crucial role since it preserves the software quality after deployment~\cite{36}. Debugging tools are used in this phase of the software development life cycle to help identify where the software faults are and what is the condition that leads to them.

The increasing need for production efficiency and flexibility require active and real-time maintenance from skilled technicians to reduce machine downtime. For expensive machines, usually, often downtime is more expensive than the actual repair in terms of lost production resource. Thus, real-time monitoring, advanced maintenance, and debugging systems play a significant role in solving the problem remotely~\cite{36}.

Cyber-physical systems are similar to traditional distributed embedded systems, consisting of several interconnected devices with limited resource constraints. However, its main goal is to remain responsive to environmental changes and network commands. They \emph{``are distributed applications that track, observe and analyze large collections of data from computerized entities''}~\cite{37}. Debugging these distributed systems is hard because developers have to deal with the non-determinism of concurrent processes, the time-sensitive nature of applications and partial failures that may occur~\cite{37}.

Bugs in CPS systems are hard to reproduce and thus to fix, and for this reason, remote debugging techniques are helpful in the maintenance process because the debugger is connected to the device when an exception or crash occurs~\cite{37}. Capturing information from these devices in real-time helps to identify and track where the causes of failure are situated.

%In the industrial manufacturing environment, the use of graphical input aids and a display interface can be used for visualizing the control of the workflow of a specific machine [SP07]. A typical workflow process is made up of a series of tasks and events, the order in which they must occur, and the script that is executed for each event.

%Because developing a workflow is usually a complex process involving many components in the system, running the workflow with a debugger to help identify and fix any issues is crucial [SBP+06]. In the ideal environment, in the context of a manufacturing system, remote debugging should be available for the debug of the workflow of a machine, with a user interface that allows a certain level of abstraction regarding the control of automation functions, either it is for controlling an industrial process programmable controller (PLC) or for programming the motion controller of a processing or production machine [SP07].

\section{Related Work}
\label{sec:relatedwork}
%----------------------------------------------------------------------------

We now present the current approaches on remote debugging, debugging of workflows and remote debugging protocols, while simultaneously identifying the shortcomings of the current approaches \emph{w.r.t.} remote debugging of IIoT workflows.

\subsection{Remote Debugging} %---------------------------------------------------------------------

JPDA~\cite{38} (Java Platform Debugger Architecture) is a debugging framework stack consisting of a mirror interface (JDI), a communication protocol (JDWP) and the debugging support running on the target virtual machine to debug (JVM TI). It splits the debugging process into the program which is being debugged and the user interface of the debugger application (JDI), communicating using the JDWP protocol through a communication channel set between the two processes, thus supporting the remote debug process by running the debug process in a different machine from the debugger.

TOD~\cite{39} is an omniscient debugger solution that enables navigation backward in time within a program execution trace. It records the events that occur during the execution of the program to be debugged and lets the user conveniently navigate through the obtained execution trace and evaluate what may have possibly gone wrong. 

The .NET debugging solution~\cite{40} provides, other than the regular debugging environment, a snapshot debugger which targets live ASP.NET apps in Azure App Service. It takes a snapshot of the in-production apps when the executing code reaches an established point, through \emph{snappoints} and \emph{logpoints}, capturing the state of the execution at that particular moment. It allows seeing what went wrong with the executing application without impacting the traffic of the application, dramatically reducing the time it takes to solve issues that occur in production environments. Both this functionality and the regular debugger support debugging remote environments~\cite{41}.

Rivet~\cite{42} is an experimental framework solution which provides browser-agnostic remote debugging of client-side web applications. It allows developers to  \emph{``inspect and modify the state of live web pages that are running inside unmodified end-user web browsers''}, allowing to explore real application bugs in the context of the actual machine in which those bugs occur. The communication between the client and the server-side is made using standard HTTP requests.

ELIoT~\cite{43} (ErLang for the Internet of Things) is a development platform for smart devices connected through the network that \emph{``provides an abstraction over the hardware on which the applications will be executed''}. It grants remote communication with the applications running on the IoT devices, having the ability to query sensors or sending a command to an actuator using CoAP services~\cite{44}.

\subsection{Workflow Debugging} %-------------------------------------------------------------------

RTWM~\cite{45} (Real-Time Workflow Monitor) provides \emph{``a cloud service and a client library to monitor complex workflow systems in real-time''}. It claims to provide the progress of the execution of each task of the workflow in real-time, monitoring the execution order and fault occurrences of the system remotely and presenting it asynchronously through a user interface. The interface provides an infrastructure for setting up the workflow system, monitoring performance and debugging a chained set of tasks. Here, a web server is running the main application for debugging and receiving events from the computing units through WebSocket connections, which then provides visualization for monitoring on the web app. The UI Layer uses HTTP and WebSocket connections for communication with the device's lower layers in real-time.

Stampede~\cite{46} (Synthesized Tools for Archiving, Monitoring Performance and Enhanced Debugging) \emph{``intends to apply an offline workflow log analysis capability to address reliability and performance problems for large, complex scientific workflows''}. It works as an execution information capture and analysis system, streaming and storing information about the performance of workflows in real-time on a log file written during execution. It also supports distributed environments. The resulting log file contains the status of each job, inputs, and outputs of tasks, and the pre and post-execute scripts.

Node-RED~\cite{47} is a programming tool for creating flow-based programs that connect hardware devices, APIs and online services. It provides a browser-based graphical editor for building the workflows, wiring the nodes that represent each task. This tool can be run locally, on a physical device or in the cloud, allowing to visually represent the flow of tasks with an abstraction level that provides the user with the ability to break down a problem into more manageable steps and look at what each produces, without having to understand the individual lines of code within each node.

\subsection{Remote Debugging Protocols} %-----------------------------------------------------------

The GDB RSP~\cite{48} (Remote Serial Protocol) is a high-level protocol that allows GDB to connect through a serial port remotely. After the communication setup process is finalized, the debugger can use the usual commands to examine and change data and to stop or continue the remote program's execution. This protocol supports a wide range of connection types.

The JDWP~\cite{49} (Java Debug Wire Protocol) is the protocol used for communication between a debugger and the JVM where the program is running. It does not specify the transportation of the messages, only details the format and the layout, and it accepts any transport mechanism that is suitable for the target debugger/target VM combination through a simple API. The packets of the messages exchanged are defined as ``command'' or ``reply'' packets. The first can be used by the VM to request information about the application's internal variables, to control the program's execution, or by the running application to notify the debugger of some event such as the reach of a breakpoint or an exception. Reply packets are sent in response to a command packet, providing information regarding the success or failure of the command to what it is replying, carrying some data if requested.

\subsection{Current Issues} %-----------------------------------------------------------------------

As presented in the above section, current solutions offer some alternatives for the debug of workflows and for remote debugging. However, none of them seems complete as a whole for these two combined and contextualized in the manufacturing industry. Similarly, there is a lack of proper testing tools for Internet-of-Things based systems~\cite{8411738}. To ensure the maintenance process for the manufacturing production line, remote control and monitoring of machines should be possible at any time, allowing to detect failures as soon as they happen (cf. \textsc{Device Error Data Supervisor}~\cite{dias17}), thus avoid any resource losses resulting from an unscheduled stop of a machine on the shop-floor.

To allow a better understanding of the information obtained through the real-time monitoring of equipment feedback, the use of visual metaphors as abstraction mechanisms seems suitable, since such approaches have been around the manufacturing and automation industry for a long time~\cite{57}. Visual programming is simply a formalization of the workflow of the tasks to be carried by the operational system. A typical workflow process is made up of a series of tasks and events, the order in which they must occur, and the script that is executed for each event~\cite{3,45}. A communication protocol is required to connect the visualization of the workflow to be monitored to the physical device where it is running. None of the solutions presented fully answer these demands simultaneously:
\begin{itemize}
\item Remote connection to a specific physical device running elsewhere without compromising its current execution;
\item The abstraction of the physical device's sequenced execution tasks through a workflow;
\item Ability to debug the workflow being executed by the physical device in real-time (with or without interrupting its execution).
\end{itemize}

Thus, we consider that there is a need for some solution that address these loose gaps to improve IIoT systems maintenance.

%-mes - falar dos workflows
%-architecture
%-solution overview

\section{Problem Statement} %-----------------------------------------------------------------------
\label{sec:problemStatement}

This section presents the solution proposed by this article. 

\subsection{Context} %------------------------------------------------------------------------------

The proposed solution was developed in the context of Critical Manufacturing's Manufacturing Execution System. It is a software platform with a deep set of modular, fully interoperable, applications which claims to \emph{``provide manufacturers in complex industries with the maximum agility, visibility, and reliability''} of the manufacturing process~\cite{58}. 

The recent advances in the integration of IIoT led to the development of the Connect-IoT module. This platform is a low-code solution that enables production engineers and system integrators to connect their shop-floor equipment to Critical Manufacturing MES. It has the goal to dramatically reduce the time and effort for this very integration, allowing to create a graphical overview of the automation workflows being carried out by the machines in production.

It has a single graphical view of automation workflows, providing ways to create and update complex logic via a user-friendly, no-code interface that allows workers with close to none IT-knowledge to fulfil their needs~\cite{59}. So far, Connect-IoT allows to visually debug these workflows and simulate the equipment, with inputs and outputs provided by the user, as the IoT application is running on the background in the same machine as the debugger for the testing of this manufacturing environment before deploying it in the production line.

Machinery downtime should be avoided at all cost since it causes dreadful resource losses for companies. To prevent this and to assure maximum production efficiency and effectiveness, real-time remote maintenance is required, which can be achieved through remote debugging of the operations' workflow being followed by the machines in the production line. We implement it as an extension to the currently available module in Critical Manufacturing MES, by receiving the workflow inputs and outputs directly from the equipment.

\subsection{Architecture}
\label{ssec:4b}
%-------------------------------------------------------------------------

\begin{figure}[htbp]
\centerline{\includegraphics[width=0.5\textwidth]{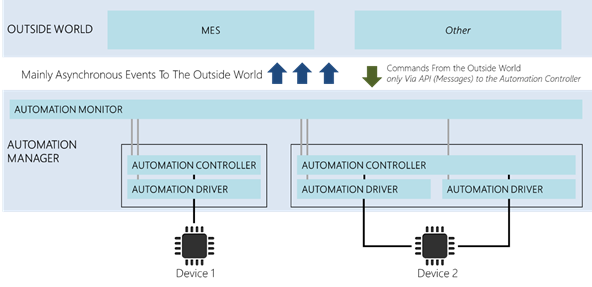}}
\caption{IoMT Architecture.}
\label{fig:architecture}
\end{figure}

To set up the environment necessary for the debug of a workflow, it's necessary to define and relate the components presented in Table~\ref{tab:iomt-components} in the Critical Manufacturing MES, which will be defined through the MES interface. The Automation Driver, Controller, and Monitor processes will be run in a computer physically close to the physical device to debug. This three-component environment will be from now on called IoMT (Internet of Manufacturing Things) agent, for easier reference (see Fig.~\ref{fig:architecture}).

\begin{table*}[htbp]
\scriptsize
\caption{IoMT components}
\begin{center}
\begin{tabularx}{\linewidth}{|p{2.3cm}|p{14.95cm}|}
\hline
\textbf{Component} & \textbf{Description} \\
\hline
Automation Protocol & 
Definition of the communication protocol to be used between the Automation Driver and the physical device (machine), such as SECS/GEM, OPC, AMQP, MQTT, HTTPS, etc. The OPC UA protocol was chosen to be used for the testing of the developed prototype. It will be associated with the Automation Driver process. \\
\hline
Automation Driver & 
A low-level process that communicates physically with the device using the defined protocol, using it to interpret the messages received and report the events from the device to the Automation Controller associated with this driver. In the definition of this component, the protocol to be used, the properties/variables of the physical device we want to monitor, the events to be detected and the commands for the Automation Controller to execute will be designated. \\
\hline
Automation Controller & 
An event-driven process that executes workflows in response to events received from the Automation Driver or directly from the MES. It can communicate with more than one driver. The definition associates the Automation Driver(s) to the device/resource we want to monitor and defines the workflows to be executed in response to the received events. When a controller is initialized, it originates an Automation Controller instance which will have a unique ID. \\
\hline
Automation Monitor & 
A process that determines which processes must be started or stopped, monitoring the health of these processes. It connects the Automation Controller and Driver(s), telling the driver processes where to find the controller it has been associated with so that they can begin communication. \\
\hline
Automation Manager & 
A process that will host one Automation Monitor and several Automation Controllers and Drivers as they were configured in the MES interface. It is responsible for spawning and controlling these processes. \\
\hline
\end{tabularx}
\label{tab:iomt-components}
\end{center}
\end{table*}

The automation workflow graphs are defined in the Automation Controller workflow designer interface. These are composed of several tasks, linked through wires that connect outputs from one workflow task to the inputs of other tasks, enabling the passing of values between tasks. These links may have value converters associated, which may change the type of the variable (useful when the variable type doesn't match with what the input is expecting) or have some small business logic to apply before it gets to the input. The workflow will be executed whenever a particular event occurs, activating it (e.g. upon initialization, when a certain device state change or upon receiving an input from the previous task) and will be executed by the controller's workflow engine during runtime.

\subsection{Solution Overview} %--------------------------------------------------------------------

%This work delves into the challenges of debugging IIoT systems, namely remote debugging strategies. A prototype to demonstrate the feasibility of both synchronized and snapshot remote debugging applied to an IIoT workflow shall be designed, constructed, and tested/evaluated. 

The main goal of this approach is to achieve is a 0-downtime workflow remote debug approach, contemplating the design and implementation of a suitable debugging protocol and a supporting platform where workflows can be easily configurable, monitored and debugged while developing or during production, choosing to interrupt or not the workflow's operations while debugging.

It will allow factory workers to debug a working machine through the network to access the feedback provided by the sensor devices and control the execution through the setup of breakpoints in the machine's workflow. This will provide access to the input and output values of each task, allowing to detect what may be causing incorrect behavior, resulting in an erroneous service in the production line. The proposed solution includes four debug modes:

\begin{itemize}
\item Mock Debug: All the workflow inputs are manually inserted into the UI of the debugger application. The IoT application runs in the same machine as the debugger just for simulating the execution of the real manufacturing environment;
\item Remote Synchronous Debug: The workflow inputs are received directly from the manufacturing equipment and/or from MES. When the execution of the workflow stops on a breakpoint, the entire workflow stops and the developer can check and modify the state of the internal variables, before proceeding with the execution;
\item Remote Snapshot Debug: Similar to the previous approach, but in this case the workflow developer can only check the current state of the workflow execution, not interrupting the source system;
\item Remote Profiler Debug: Same as the Snapshot Debug mode approach, changing only in regards to the way the information is presented to the user.
\end{itemize}

\subsection{Desiderata} %-------------------------------------------------------------------

The following describe the features that we aim to tackle in this work:

\begin{itemize}

\item There should be an adaptation of a debug session to the manufacturing environment, where there are events defined to be triggered by a certain change in the machine itself which will then have to be communicated to the debugger.

\item The remote debugging protocol should be generalized so that the number of messages is reduced and can be used for the different debug modes. Each message should have the purpose of informing or notifying the receiver of something that happened or must happen and must be related to a distinct action.

\item The debugger should be aware of all the debug sessions it is a participant of, and there should be a frequent checkup for inconsistencies, using the debugging protocol, regarding sessions that remain available, the debug mode type of each session, etc., according to the availability rules to be defined in the section below, so that the machine's execution is not being interrupted when it shouldn't.

\item The remote debugging method should provide a better insight of the production machinery, even if it is in execution (through the snapshot debug mode), in a real-life production line. 

\item  Creating an abstraction of the execution tasks using a representation through workflows should provide a better understanding of the machine execution and allow to detect failures in real-time and what is causing the incorrect behavior, avoiding losses resulting from machinery downtime. This can be achieved through a simple interface that allows knowing what inputs were received and what outputs were generated by a task and the variable's evolution through the workflow path, presenting the debug information intuitively so that the general factory worker with close to none IT knowledge can operate with it.

\end{itemize}

\section{Proposed Solution}
\label{sec:proposedSolution}

%This section details the proposed solution, explaining its main software components and the techniques used to tackle the development problems. 

Our approach focuses on updating the existing components of the MES software to support the new functionalities, more specifically, the Automation Controller process, mentioned in the ``Architecture'' subsection, and the Automation Controller view on MES where it is possible to build the workflows and debug them locally.

Because it is the Automation Controller process that will be running the workflow, receiving the events from the Automation Driver(s) connected to the device we want to debug, the communication between this component and the MES instance is essential to allow the remote debug functionalities that this work aims to implement. For this, the modified components have to behave accordingly with the definition of the remote debugging protocol developed.

After the environment setup is done on MES, in the Automation Controller workflow view of the MES interface, a dropdown list was added to allow the selection of the debug mode: (1) Mock Debug, (2) Synchronous Remote Debug, (3) Snapshot Remote Debug and (4) Profiler Remote Debug.

%\begin{figure}[htbp]
%\centerline{\includegraphics[width=0.5\textwidth]{images/1.png}}
%\caption{Debug mode selection.}
%\label{fig:1}
%\end{figure}

The Mock Debug mode debugs an IoT application running in the same machine as the debugger using workflow inputs manually inserted into the UI of the debugger application. The workflow engine is running locally. This is useful for simulating the execution of a real manufacturing environment.

The remaining debug modes in the list interact with an IoT application running remotely, in physical devices executing in a real production factory environment, receiving the workflow inputs directly from the manufacturing equipment. 

Because these modes require a connection with the remote entities (the IoMT agent), a communication protocol was developed. To sent the messages between these entities, the message bus that was already implemented in the Critical Manufacturing MES, which uses WebSockets, was used. This way, all messages will go directly to the message bus gateway, being diffused by all the MES instances and by the IoMT agents running the Automation Controller instances close to the machines. Four of the implemented methods in the message bus were used by this communication protocol: publish (sends a message asynchronously), sendRequest (sends a message synchronously and waits for the response), reply (replies to a sendRequest message), subscribe (subscribes to a message of a certain subject and registers the callback to be called when a message of this type is received).

In the remote debugging protocol, each message identifies the components it wants to be associated with as attachments to the message subject. An overview can be found in Table~\ref{tab:protocol}.

\begin{table*}[!htbp]
\scriptsize
\begin{center}
\caption{Remote Debugging Protocol}
\begin{tabularx}{\textwidth}{|p{3.2cm}|p{1.2cm}|p{0.62cm}|p{5cm}|p{2.2cm}|p{3.3cm}|}

\hline

\textbf{Message} &  \textbf{Flow} & \textbf{Type} & \textbf{Description} & \textbf{Parameters} & \textbf{Reply}   \\

\hline

\hskip 0pt plus 0pt onCommunicationStarted & \hskip 0pt plus 0pt IoMT-MES & Async & 
Connection attempt from an IoMT agent upon initialization with all the MES instances available. It is sent to start/restart the communication between them and inform if the workflow instance selected on the debugger page is running in that server.
& \hskip 0pt plus 0pt automationControllerInstanceId: string & \hskip 0pt plus 0pt onCheckWorkflowRunning- \_(automationControllerInstanceId) \\

\hline

\hskip 0pt plus 0pt onCommunicationAttempt- \_(automationControllerInstanceId)
& \hskip 0pt plus 0pt MES-IoMT & \hskip 0pt plus 0pt Sync & 
Connection attempt from a MES instance upon selection of the Automation Controller instance to debug, repeated every 10 seconds to ensure it detects a change in the connection if it happens.
& \hskip 0pt plus 0pt reply: function & \hskip 0pt plus 0pt reply \\

\hline

\hskip 0pt plus 0pt onCheckWorkflowRunning- \_(automationControllerInstanceId)
& \hskip 0pt plus 0pt MES-IoMT & Async / Sync & 
Communicates with the selected Automation Controller instance to know if the workflow instance selected on the debugger page is running in that server. The answer is received in the MES instance through the onCheckWorkflowRunningResponse\_(auto- mationControllerInstanceId)\_(workflowId) message (async) or through a reply in the start debug function (sync). This will get a reply if it is running or get no reply otherwise.
& \hskip 0pt plus 0pt workflowId: string, sessionId: string, reply?: function
 & \hskip 0pt plus 0pt onCheckWorkflowRunning- Response\_(automationCon- trollerInstanceId)\_(work- flowId), reply \\
 
\hline

\hskip 0pt plus 0pt onCheckWorkflowRunning- Response\_(automationCon- trollerInstanceId)\_(work- flowId)
& \hskip 0pt plus 0pt IoMT-MES & Async & 
Checks if, according to the sessionIds related with the selected workflow instance, the workflow is available for debugging or not.
& \hskip 0pt plus 0pt running: boolean, sessionsId: DebugSessionInfo[] & \\

\hline

\hskip 0pt plus 0pt onBreakpointChange\_- (automationControllerIn- stanceId)\_(workflowId)
& \hskip 0pt plus 0pt MES-IoMT & Async & 
Adds or removes a breakpoint on the selected workflow instance.
& \hskip 0pt plus 0pt sessionId: string, breakpoint: BreakpointDefinition & \\

\hline

\hskip 0pt plus 0pt onBreakpointToggle\_- (automationControllerIn- stanceId)\_(workflowId)
& \hskip 0pt plus 0pt MES-IoMT & Async & 
Enables or disables a breakpoint on the selected workflow instance.
& \hskip 0pt plus 0pt sessionId: string, breakpoint: BreakpointDefinition & \\

\hline

\hskip 0pt plus 0pt onStartDebug\_(automation- ControllerInstanceId)
& \hskip 0pt plus 0pt MES-IoMT & Async & 
Initializes a debug session in the selected Automation Controller instance, if available, generating a unique debug session Id and updating the workflow breakpoints associated with the session.
& \hskip 0pt plus 0pt mesId: string, workflowId: string, debugMode: DebugMode, breakpoints: BreakpointDefinition[] & \hskip  0pt plus 0pt onDebugStarted\_(mesId)\_- (automationControllerIn- stanceId)\_(workflowId) \\

\hline

\hskip 0pt plus 0pt onDebugStarted\_(mesId)\_- (automationControllerIn- stanceId)\_(workflowId)
& \hskip 0pt plus 0pt IoMT-MES & Async & 
Communicates the newly created debug session Id to the MES instance that asked for the start of that debug session.
& \hskip  0pt plus 0pt sessionId: string & \\

\hline

\hskip 0pt plus 0pt onStopDebug\_(automation- ControllerInstanceId)
& \hskip 0pt plus 0pt MES-IoMT & Async & 
Stops an existing debug session.
& \hskip  0pt plus 0pt sessionId: string & \hskip  0pt plus 0pt onDebugStopped\_(automa- tionControllerInstanceId)\_- (sessionId) \\

\hline

\hskip 0pt plus 0pt onDebugStopped\_(automa- tionControllerInstanceId)\_- (sessionId)
& \hskip 0pt plus 0pt IoMT-MES & Async & 
Updates the session registry array.
& \hskip  0pt plus 0pt sessionId: string, registry: DebugSessionInfoEntry[] & \\

\hline

\hskip 0pt plus 0pt onSessionRenewal\_(auto- mationControllerInstanceId)
& \hskip 0pt plus 0pt MES-IoMT & Async & 
Renews an existing debug session. This is necessary to guarantee that the session hasn't expired and remains active over time.
& \hskip  0pt plus 0pt sessionId: string &  \\

\hline

\hskip 0pt plus 0pt onBeforeSetOutputs\_(automa- tionControllerInstanceId)\_- (sessionId)
& \hskip 0pt plus 0pt IoMT-MES & Async / Sync & 
Notifies the MES instance debugging a certain workflow that a variable on a task output with a breakpoint has been changed, changing the visual representation of the workflow in the interface.
& \hskip  0pt plus 0pt workflowId: string, registryEntry: DebugSessionInfoEntry, reply?: function & 
\hskip  0pt plus 0pt reply \\

\hline

\hskip 0pt plus 0pt onAfterSetInputs\_(automa- tionControllerInstanceId)\_- (sessionId)
& \hskip 0pt plus 0pt IoMT-MES & Async / Sync & 
Notifies the MES instance debugging a certain workflow that a variable on a task input with a breakpoint has been changed, changing the visual representation of the workflow in the interface.
& \hskip  0pt plus 0pt workflowId: string, registryEntry: DebugSessionInfoEntry, reply?: function & 
\hskip  0pt plus 0pt reply \\

\hline

\hskip 0pt plus 0pt onReceivedExecutionCon- text\_(automationController- InstanceId)\_(workflowId)
& \hskip 0pt plus 0pt MES-IoMT & Async & 
Saves the execution context of an snapshot debug session it wants to follow.
& \hskip  0pt plus 0pt sessionId: string, executionContext: string &  \\

\hline

\hskip 0pt plus 0pt onAvailableACIRequest
& \hskip 0pt plus 0pt MES-IoMT & Async & 
Requests an update of all running Automation Controller instances of the current running state of the selected workflow in the debugger page.
& \hskip  0pt plus 0pt workflowId: string & \hskip  0pt plus 0pt onAvailableACIRequest- Response\_(workflowId) \\

\hline

\hskip 0pt plus 0pt onAvailableACIRequest- Response\_(workflowId)
& \hskip 0pt plus 0pt IoMT-MES & Async & 
Updates the currently available Automation Controller instances array with the information received regarding the Automation Controller instance.
& \hskip  0pt plus 0pt automationControllerInstanceId: string, running: boolean &  \\

\hline

\end{tabularx}
\label{tab:protocol}
\end{center}
\end{table*}

\subsection{Connection Preparation}

When reaching the debugger page, both the debug mode and workflow to debug are already selected so, from there on, when remaining in this page, all the communication will be regarding that workflow and will have in consideration the debug mode that was selected. 
The only element to select that remains missing is the Automation Controller instance we wish to connect with to begin the debug of the workflow. For that, the \texttt{onAvailableACIRequest} message is broadcasted to all the entities connected to the message bus, sending the ID of the selected workflow. The Automation Controller instances subscribed to this message will reply with the message \texttt{onAvailableACIRequestResponse} informing their unique ID and the execution state of the workflow in that controller instance. According to the response received, the MES instance will gather the IDs of the Automation Controller instances where the selected workflow is running. If there is more than one positive response, the IDs will be displayed in a dialog for the user to select which controller instance they wish to debug. If only one controller instance replied within 5 seconds, the MES instance will assume it is the only one running the workflow and will automatically select it.

Once the Automation Controller instance selection is over, the communication between this instance and MES begins, ensuring the connection between these over time until the user stops the communication (for example, by leaving the debugger page). Once the controller instance selection is done, the MES instance sends an \texttt{onCheckWorkflowRunning} to know if the workflow is still running and if it is available for debugging in the selected mode. This message is more refined than the previous one used to know if the workflow is running because it will also receive the debug sessions (and respective mode type) associated with it, determining its availability: 
\begin{itemize}
\item Only one synchronous debug session can be active at once, and no other debug sessions are allowed until that session has ended (synchronous or snapshot);
\item There can be more than one snapshot session at the same time, since they don't interfere with the machine's execution, but no synchronous session can start until all snapshot sessions have ended.
\end{itemize}

The response, \texttt{onCheckWorkflowRunningResponse}, will determine, according to these rules and the chosen debug session, if the workflow is available for debugging or not, determining the state of the ``Start Debug'' button (enabled/disabled).

To keep the consistency of the information presented, both \texttt{onAvailableACIRequest} and \texttt{onCommunicationAttempt} requests are sent periodically, with the intervals of 30 and 10 seconds respectively.  In case the \texttt{onCommunicationAttempt} message got a positive response, it will lead to the dispatch of the \texttt{onCheckWorkflowRunning} request, along with the message to renew the debug session state, if it has one, \texttt{onSessionRenewal}. 

The session renewal state ensures that the sessions that unexpectedly disconnect over time are wiped by the IoMT agents periodically (every 30 seconds). The same is done regarding the available controller instances, which have a renew state that wipes the ones that unexpectedly stopped replying to the \texttt{onAvailableACIRequest} messages, every 30 seconds.
 If for some reason, the connection between the two entities (MES and  IoMT) drops because the selected IoMT agent has disconnected, it will send an \texttt{onCommunicationStarted} message upon initialization in case it was connected to a MES instance to quickly re-establish the connection.

\subsection{Debug Session}

To start the debugging session, when the ``Start Debug'' button is clicked, the \texttt{onStartDebug} request is sent to the selected Automation Controller instance. It will re-check all the conditions previously mentioned to know if the workflow is available for debugging and if so, it will generate a unique session ID and update the workflow breakpoints associated with the debug session. The session ID will be communicated to the respective MES instance through the \texttt{onDebugStarted} message, using the unique MES instance identifier.

In the Automation Controller, during the equipment setup, the engine that is running the workflow will associate the Automation Driver to each task so that when a variable change happens, the driver will detect the changes and will emit an event subscribed by the respective task input/output (the subscription is done during the engine's workflow setup). When this event is detected by the task, it will trigger the \texttt{onBeforeSetOutputs} or the \texttt{onAfterSetInputs} hook respectively in case it was a task output or input. This process is common to all debug modes (remote or not).

In the case of the Mock Debug, the workflow engine is running next to the MES instance, and the hooks behave slightly differently. The functions associated with the hooks are directly injected in the local engine, detecting the modifications in the task's inputs or outputs, which will then trigger the breakpoint, if there is any, in the modified variable, stopping the workflow execution when a breakpoint is reached until the ``Resume'' button is pressed.

When working with a remote Automation Controller instance, when these hooks are triggered, it will identify the debug sessions associated with the workflow where the changes were detected and, for each, it will check if there are any breakpoints in the modified variable.
    
In the case of the Synchronous Debug, if there is a breakpoint in that variable, it will stop the engine's execution, send an \texttt{onAfterSetInputs} or \texttt{onBeforeSetOutputs} message, respectively in the case of an input or output, notifying the MES instance of the workflow state when the breakpoint was reached, the variable changes and the triggered breakpoint. The Automation Controller instance will then wait until it receives a reply to these messages.

When receiving the previously mentioned messages, the MES instance will trigger the respective breakpoint, update the workflow task instances to show the workflow state when the changes were detected and stop the engine's execution until the ``Resume'' button is pressed, which will then resume execution of both MES and Automation Controller instance workflow engines by replying to these messages.

Because when the engine is stopped the hooks will still get triggered when a variable is changed, in this mode, all the hooks triggered while waiting for the ``Resume'' command will be ignored by the engine, along with the propagation of these new variables in the workflow.

In the case of the Snapshot or Profiler Debug, if there is a breakpoint in the modified variable, it will check the execution context where this change was detected, and update the registry of changes on that execution context.

This introduces the concept of a ``flow'' of the workflow, which begins when an equipment event is triggered and ends at the end of the workflow branch where there is no output link. Each flow will have its own execution context and will not interfere with the execution of another flow.

This was a mechanism made for dealing with asynchronous workflows where, for example, two variables will trigger the same output, but at different times. Imagining that two variables will change consecutively, and the concept of flows hasn't been implemented, and variable A will change prior to variable B but will take longer to reach the common output. The presented information to the user will be $ var\,A \rightarrow var\,B \rightarrow output\,B \rightarrow output\,A $. Assuming that the output will be the result of calculations with the received variables but the user does not have insight of those calculations, it will assume the output with variable B was actually gotten using the variable A, since that one was changed first. With the introductions of flows, the user will be presented with only one workflow flow and the information of different execution contexts will not get mixed, making the workflow much easier to debug. Whenever an equipment event is triggered by a variable change, it creates a new execution context which will follow the evolution of that variable through the workflow.

The Snapshot and Profiler Debug modes work similarly, changing only in regards to the way they present the information to the user.  

In the case of the Snapshot Debug, we only want to follow the most recent detected execution context. One snapshot session debugs only one execution context to understand exactly what was the evolution of a variable through the workflow and the path it took. When a variable change is detected, if there is a breakpoint in this variable, the Automation Controller instance will send an \texttt{onAfterSetInputs} or \texttt{onBeforeSetOutputs} message, respectively in the case of an input or output, notifying the MES instance of the workflow state when the breakpoint was reached, the variable changes and the triggered breakpoint. This will not affect the execution of the engine, and the workflow will keep triggering the event hooks, communicating the respective messages.

When receiving these messages, the MES instance will trigger the received breakpoint, update the workflow task instances to show the workflow state when the changes were detected and stop in that breakpoint until the ``Resume'' button is pressed. Because the engine's execution is not altered, when receiving a message while waiting at a breakpoint, it will stack the breakpoint promises. So when the ``Resume'' button is pressed, it will go to the next triggered breakpoint, if there is any, or wait for the next message.

Upon receiving the first variable change message, it will check the execution context in which this event was triggered and will send the \texttt{onReceivedExecutionContext} request to inform the Automation Controller instance that it wishes to follow that execution context. This will act as a filter in the event hooks to know which messages should be sent to the MES instances and ignore all variable changes with a different execution context as the one that was chosen for this debug session.

\begin{figure*}[!htb]
\centering
\includegraphics[width=1\textwidth]{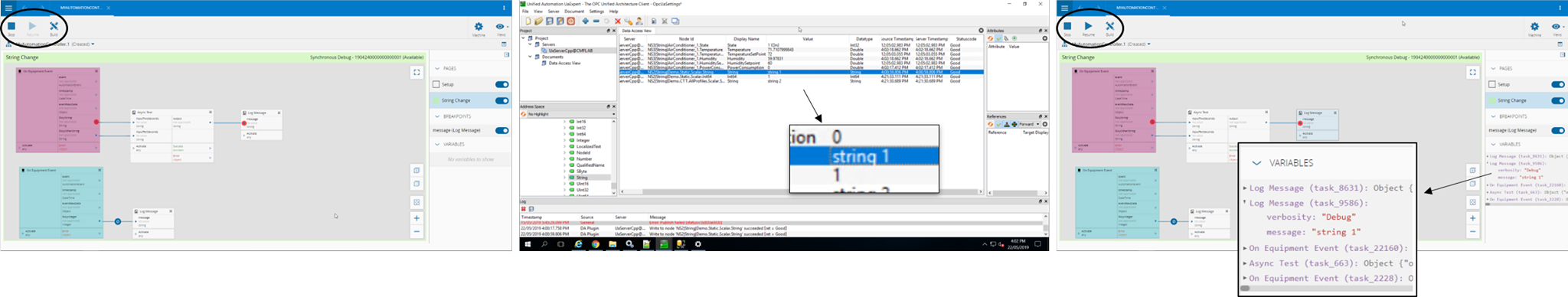}
\caption{Synchronous Debug session.}
\label{fig:6}
\end{figure*}

\begin{figure*}[!htb]
\centering
\includegraphics[width=1\textwidth]{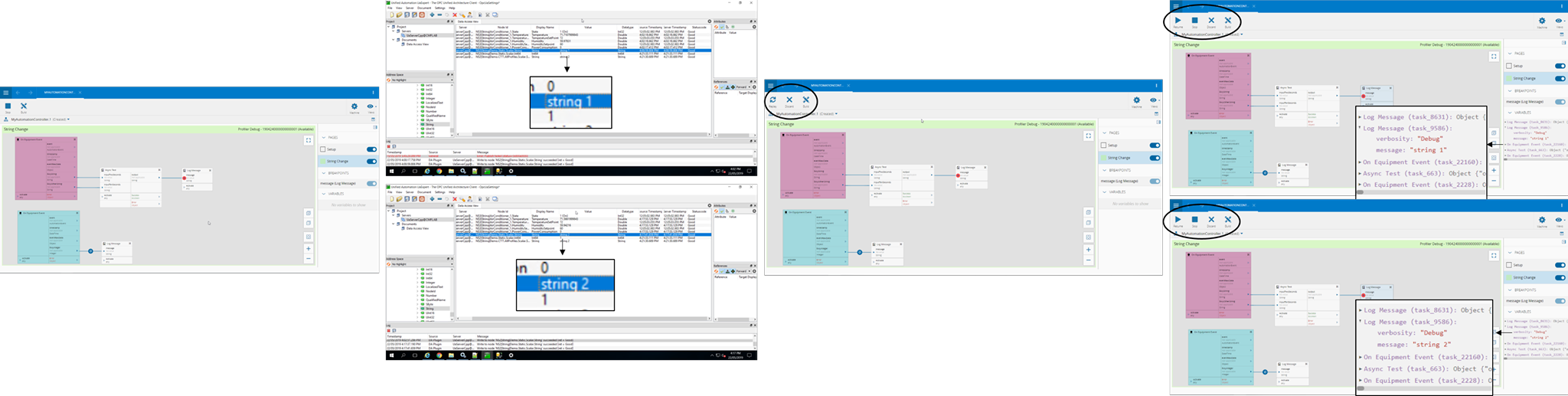}
\caption{Profiler Debug session.}
\label{fig:7}
\end{figure*}

\begin{table*}[!h]
\scriptsize
\begin{center}
\caption{Simulated Scenarios and Results}
\begin{tabularx}{\linewidth}{|p{7.4cm}|p{9.9cm}|}

\hline

\textbf{Scenario} &  \textbf{Expected Result}  \\

\hline

MES instance receives information about the availability of the selected workflow on an Automation Controller & If available, the Id of the Automation Controller instance will show up in the Automation Controller list for selection.\\

\hline

MES instance connects to the selected Automation Controller instance & Id of the selected Automation Controller shows up in the ribbon on top of the workflow in the MES instance interface, along with the current availability of the workflow.\\

\hline

MES instance can handle a disconnect, followed by a reconnect, of the selected Automation Controller instance & Connect/Disconnect messages will show up and the state of both workflow availability and "Start Debug" button will change accordingly. Will stop any active debug session upon disconnect. \\

\hline

MES instance can begin a synchronous debug session and show the variable changes as they are received & When there is an equipment variable change and a breakpoint is linked to that variable, the Automation Controller gets a notification and transmits it to the MES instance. This activates the breakpoint in the interface and shows the state of the workflow, at the moment the breakpoint was triggered, until the ``Resume'' button is pressed (Fig.~\ref{fig:6}). \\

\hline

Automation Controller instance ignores any variable change when waiting for a response from the MES instance during a synchronous debug session & Nothing happens in the MES instance whilst it is waiting at a breakpoint. \\

\hline

MES instance can begin a snapshot debug session and show the variable changes as they are received, stacking the breakpoints received if required & Same as the synchronous debug, but if there are any changes whilst waiting for a ``resume'' response, the breakpoint promises stack and the next breakpoint will activate when the ``Resume'' button for the current breakpoint is pressed. \\

\hline

MES instance doesn't show any variable change during a snapshot debug session if it is not from the chosen execution context & Variable changes from other execution contexts are ignored and nothing happens in the MES instance relative to those. \\

\hline

 MES instance can begin a profiler debug session and show the variable changes collected by the Automation Controller instance, chronologically, through replay mode & Collects variable changes from all execution contexts during the debug session. When stopped, sends the information to the MES instance where the user will see all changes that happened, chronologically, whilst debugging, using the breakpoint activation method used in the snapshot debugging (Fig.~\ref{fig:7}). \\

\hline

MES instance doesn't let any other debug session start if there is already a synchronous debug session for the selected workflow & ``Start Button'' of all other debug sessions is disabled if there is already a synchronous debug session. \\

\hline

MES instance doesn't let any synchronous debug session start if there is at least one snapshot debug session for the selected workflow & "Start Button" of all synchronous debug sessions is disabled if there is at least one snapshot debug session. \\

\hline

MES instance allows changing breakpoints during a synchronous session, and they are immediately updated in the Automation Controller instance & Clicking in the workflow input/outputs adds/removes breakpoints. \\

\hline

MES instance doesn't allow changing breakpoints during a snapshot/profiler session & Clicking on a breakpoint whilst in an snapshot or profiler session won't do anything. \\

\hline

\end{tabularx}
\label{tab:scenarios}
\end{center}
\end{table*}

In the case of the Profiler Debug, it will collect the information of the different execution contexts passively, without notifying the user of the changes that are happening in the machine, until the session is stopped (cf. \textsc{Device  Raw  Data Collector}~\cite{dias17}). This will allow the user to check all the events that happened while the debug session was active, chronologically. When the session is stopped, if the debug mode selected is the Profiler mode, the Automation Controller instance will reply with the \texttt{onDebugStopped} message which will sort the registry entries of all created execution contexts chronologically since the beginning of the debug session and send them to the MES instance. The MES interface will then change to allow a ``Replay Mode'' which will go through the received registry so that the user can carefully analyze the collected data. When the registry is discarded, a new profiler debug session may begin.

To stop any remote debug session, the \texttt{onStopDebug} message is sent when the ``Stop Debug'' button is clicked. After sending the necessary information to finalize the debug session to the MES instance, it will delete all the session information kept on the IoMT agent (chosen execution context, registry, breakpoints, etc). The information kept in the MES instance is erased differently according to the debug mode: 
\begin{itemize}
    \item Synchronous: done immediately after the session is stopped;
    \item Snapshot: done after the last breakpoint promise is resumed;
    \item Profiler: done when discarding the session registry.
\end{itemize}

Regarding the workflow breakpoints, they are associated with a debug session, allowing multiple snapshot sessions to have their breakpoints independently. The breakpoints are immediately set as they are displayed in the workflow debugger page when a debug session is started. The snapshot type sessions do not allow any breakpoint changes while a session is active (or while the session registry wasn't discarded in the case of the Profiler mode) because this could cause some unexpected results when receiving the events of a breakpoint that was changed while the session was active. For the synchronous session, it is possible to change the breakpoints while the session is active. Breakpoints can be added or removed in the workflow display through the \texttt{onBreakpointChange} request or can be toggled to enabled or disabled through the \texttt{onBreakpointToggle} request.

\section{Experiments}
\label{sec:experiments}

To validate our approach, a list of simulated scenarios was made to understand how the implemented features work in the already existing environment and how the debugging modes interact with one another, using the developed protocol. Scenarios regarding the management of connection between the entities were also considered.

To test the debugger application quickly, an OPC UA server was used for simulating real-time variable changes. These servers play a crucial role as a communication gateway, allowing OPC UA clients to access HMI or PLC data by subscribing to tags to receive real-time updates~\cite{29}. For this reason, the Automation Driver with the OPC UA protocol associated was the one used for testing the application.
 
%The workflow used for simplification and to ease the understanding of the debugging functionalities developed consists of an event which will get triggered when a variables changes. Then the variables will go through an asynchronous task where one input will take 2 seconds to be passed as an output and the other 10 seconds, and then they are printed in the console through the ``Log Message'' task.

%\subsection{Simulated Scenarios and Results}

The simulated scenarios and individual results are presented in Table~\ref{tab:scenarios}. These were made in order to answer the requirements defined in Section~\ref{sec:problemStatement} and to ensure that the solution developed did encompass all test cases and everything is working as it was defined in the proposal. All of the results obtained matched the expected results.

\section{Final Remarks}
\label{sec:conclusion}

%This final section of the article presents the main contributions and conclusions of this work, ending with a description of the future work that is planned.

The main conclusion of this work is that it is possible to use an abstraction of the production machines execution tasks through workflows to remotely debug them in real-time with real data provided by the sensors and actuators of said machine. Current solutions for remote debugging do not offer an approach that is simultaneously oriented towards the manufacturing industry and supports debugging of workflows. The remote debugging protocols found did also not cover what was necessary for the fulfilment of the defined requirements. The solution presented in this document solves this through the development of a protocol aimed at the desired objective, proving the extensive functionality of such solution through the experiments conducted in the ``Evaluation'' section.

The main contributions of this work are focused on the implementation of IoT in the manufacturing industry and how that can help minimize incorrect machinery behaviour through the usage of real-time debugging during the maintenance phase whilst in production:
\begin{itemize}
    \item Synchronous and snapshot debug functionalities that support the manufacturing environment;
    \item A generalized remote debugging protocol that serves both synchronous and snapshots debug particularities, taking into consideration various factors that may interfere with the correct execution of these debugging sessions;
    \item An intuitive, user-friendly way of presenting the debug information, allowing the general factory worker with close to none IT knowledge to operate with it, with the abstraction of the machine's execution tasks by using a representation through workflows. This simplified interface provides a better understanding of the machine execution and allows to detect in real-time what can be causing incorrect behaviour.
\end{itemize}

\subsection{Future Work}

As for the future work to improve the current solution, testing this solution in a real-life environment is required. The impact of its usage in the production environment should be measured to know if there is a better insight of the production machinery, as expected, and incorrect behaviour is being detected more quickly than without this debug solution.

When using the Synchronous Debug mode, it should be not only possible to see the values of the variables, but also to change them in real-time, affecting the target device being debugged. At the same time, a task's execution is stopped by a breakpoint.

As an improvement, choosing the target machine/resource to debug should also be possible using the target's IP address. The target Automation Controller identification should also be more detailed (it only uses the randomized ID it is given).

\bibliographystyle{IEEEtran}
\bibliography{refs}

\begin{thebibliography}{10}
\providecommand{\url}[1]{#1}
\csname url@samestyle\endcsname
\providecommand{\newblock}{\relax}
\providecommand{\bibinfo}[2]{#2}
\providecommand{\BIBentrySTDinterwordspacing}{\spaceskip=0pt\relax}
\providecommand{\BIBentryALTinterwordstretchfactor}{4}
\providecommand{\BIBentryALTinterwordspacing}{\spaceskip=\fontdimen2\font plus
\BIBentryALTinterwordstretchfactor\fontdimen3\font minus
  \fontdimen4\font\relax}
\providecommand{\BIBforeignlanguage}[2]{{%
\expandafter\ifx\csname l@#1\endcsname\relax
\typeout{** WARNING: IEEEtran.bst: No hyphenation pattern has been}%
\typeout{** loaded for the language `#1'. Using the pattern for}%
\typeout{** the default language instead.}%
\else
\language=\csname l@#1\endcsname
\fi
#2}}
\providecommand{\BIBdecl}{\relax}
\BIBdecl

\bibitem{54}
\BIBentryALTinterwordspacing
R.~Buyya and A.~Vahid~Dastjerdi, \emph{Internet of Things: An Overview}.\hskip
  1em plus 0.5em minus 0.4em\relax Morgan Kaufmann, 2016, pp. 3--23. [Online].
  Available:
  \url{https://www.sciencedirect.com/book/9780128053959/internet-of-things}
\BIBentrySTDinterwordspacing

\bibitem{55}
A.~Kanawaday and A.~Sane, ``Machine learning for predictive maintenance of
  industrial machines using iot sensor data,'' in \emph{8th IEEE International
  Conference on Software Engineering and Service Science (ICSESS)}, 2017,
  Conference Proceedings, pp. 87--90.

\bibitem{8411738}
J.~P. {Dias}, F.~{Couto}, A.~C.~R. {Paiva}, and H.~S. {Ferreira}, ``A brief
  overview of existing tools for testing the internet-of-things,'' in
  \emph{2018 IEEE International Conference on Software Testing, Verification
  and Validation Workshops (ICSTW)}, 2018, pp. 104--109.

\bibitem{1}
Y.~Lu, ``Industry 4.0: A survey on technologies, applications and open research
  issues,'' \emph{Journal of Industrial Information Integration}, vol.~6, pp.
  1--10, 2017.

\bibitem{2}
T.~I.~I. Consortium, ``A global nonprofit partnership of industry, government
  and academia,'' 2013.

\bibitem{3}
U.~A. Pozdnyakova, V.~V. Golikov, I.~A. Peters, and I.~A. Morozova, ``Genesis
  of the revolutionary transition to industry 4.0 in the 21st century and
  overview of previous industrial revolutions,'' in \emph{Studies in Systems,
  Decision and Control}.\hskip 1em plus 0.5em minus 0.4em\relax Springer
  International Publishing, 2019, vol. 169, pp. 11--19.

\bibitem{8}
E.~Sisinni, A.~Saifullah, S.~Han, U.~Jennehag, and M.~Gidlund, ``Industrial
  internet of things: Challenges, opportunities, and directions,'' \emph{IEEE
  Transactions on Industrial Informatics}, vol.~14, no.~11, pp. 4724--4734,
  2018.

\bibitem{4}
\BIBentryALTinterwordspacing
A.~Whitmore, A.~Agarwal, and L.~Da~Xu, ``The internet of things—a survey of
  topics and trends,'' \emph{Information Systems Frontiers}, vol.~17, no.~2,
  pp. 261--274, 2015. [Online]. Available:
  \url{https://doi.org/10.1007/s10796-014-9489-2}
\BIBentrySTDinterwordspacing

\bibitem{5}
\BIBentryALTinterwordspacing
S.~Li, L.~D. Xu, and S.~Zhao, ``The internet of things: a survey,''
  \emph{Information Systems Frontiers}, vol.~17, no.~2, pp. 243--259, 2015.
  [Online]. Available: \url{https://doi.org/10.1007/s10796-014-9492-7}
\BIBentrySTDinterwordspacing

\bibitem{6}
\BIBentryALTinterwordspacing
P.~Sethi and S.~R. Sarangi, ``Internet of things: Architectures, protocols, and
  applications,'' \emph{Journal of Electrical and Computer Engineering}, vol.
  2017, p.~25, 2017. [Online]. Available:
  \url{https://doi.org/10.1155/2017/9324035}
\BIBentrySTDinterwordspacing

\bibitem{7}
A.~Azizi, ``Modern manufacturing,'' in \emph{SpringerBriefs in Applied Sciences
  and Technology}.\hskip 1em plus 0.5em minus 0.4em\relax Springer, Singapore,
  2019, pp. 7--17.

\bibitem{9}
P.~Lade, R.~Ghosh, and S.~Srinivasan, ``Manufacturing analytics and industrial
  internet of things,'' \emph{IEEE Intelligent Systems}, vol.~32, no.~3, pp.
  74--79, 2017.

\bibitem{11}
\BIBentryALTinterwordspacing
C.~Manufacturing, ``Critical manufacturing - critical manufacturing mes -
  integrated manufacturing execution system,'' accessed: 2019-02-1. [Online].
  Available:
  \url{http://www.criticalmanufacturing.com/en/critical-manufacturing-mes/overview}
\BIBentrySTDinterwordspacing

\bibitem{12}
\BIBentryALTinterwordspacing
------, ``Critical manufacturing - what is mes?'' accessed: 2019-01-02.
  [Online]. Available:
  \url{http://www.criticalmanufacturing.com/en/critical-manufacturing-mes/what-is-manufacturing-execution-system}
\BIBentrySTDinterwordspacing

\bibitem{10}
S.~Waschull, J.~C. Wortmann, and J.~A.~C. Bokhorst, ``Manufacturing execution
  systems: The next level of automated control or of shop-floor support?'' in
  \emph{Advances in Production Management Systems. Smart Manufacturing for
  Industry 4.0}, I.~Moon, G.~M. Lee, J.~Park, D.~Kiritsis, and G.~von
  Cieminski, Eds.\hskip 1em plus 0.5em minus 0.4em\relax Springer International
  Publishing, 2018, Conference Proceedings, pp. 386--393.

\bibitem{13}
\BIBentryALTinterwordspacing
C.~Manufacturing, ``Critical manufacturing - a brief history of manufacturing
  execution systems,'' accessed: 2019-01-02. [Online]. Available:
  \url{http://www.criticalmanufacturing.com/en/newsroom/blog/posts/blog/mes-history-59}
\BIBentrySTDinterwordspacing

\bibitem{14}
\BIBentryALTinterwordspacing
V.~Roblek, M.~Meško, and A.~Krapež, ``A complex view of industry 4.0,''
  \emph{SAGE Open}, vol.~6, no.~2, p. 2158244016653987, 2016. [Online].
  Available: \url{https://doi.org/10.1177/2158244016653987}
\BIBentrySTDinterwordspacing

\bibitem{15}
\BIBentryALTinterwordspacing
M.~Naedele, H.-M. Chen, R.~Kazman, Y.~Cai, L.~Xiao, and C.~V.~A. Silva,
  ``Manufacturing execution systems: A vision for managing software
  development,'' \emph{Journal of Systems and Software}, vol. 101, pp. 59--68,
  2015. [Online]. Available:
  \url{http://www.sciencedirect.com/science/article/pii/S0164121214002532}
\BIBentrySTDinterwordspacing

\bibitem{20}
E.~R. Alphonsus and M.~O. Abdullah, ``A review on the applications of
  programmable logic controllers (plcs),'' \emph{Renewable and Sustainable
  Energy Reviews}, vol.~60, pp. 1185--1205, 2016.

\bibitem{21}
B.~Vogel-Heuser, J.~Fischer, S.~Rösch, S.~Feldmann, and S.~Ulewicz,
  ``Challenges for maintenance of plc-software and its related hardware for
  automated production systems: Selected industrial case studies,'' in
  \emph{2015 IEEE International Conference on Software Maintenance and
  Evolution (ICSME)}, 2015, Conference Proceedings, pp. 362--371.

\bibitem{22}
A.~Almohammad, J.~F. Ferreira, A.~Mendes, and P.~White, ``Reqcap: Hierarchical
  requirements modeling and test generation for industrial control systems,''
  in \emph{2017 IEEE 25th International Requirements Engineering Conference
  Workshops (REW)}, 2017, Conference Proceedings, pp. 351--358.

\bibitem{23}
I.~61131-3, ``Programmable logic controllers – part 3: Programming
  languages,'' in \emph{IEC Standard 61131-3}, 1990.

\bibitem{24}
K.~S.~K. Ibrahim, J.~H. Yahaya, Z.~Mansor, and A.~Deraman, ``Towards the
  quality factor of software maintenance process: A review,'' in \emph{Journal
  of Telecommunication, Electronic and Computer Engineering (JTEC)}, vol.~9,
  2017, pp. 115--118.

\bibitem{25}
A.~Avizienis, J.~C. Laprie, and B.~Randell, ``Fundamental concepts of
  dependability,'' in \emph{3rd IEEE Information Survivability Workshop
  (ISW-2000), Boston, Massachusetts, USA}, 10 2000, Conference Proceedings, pp.
  7--12.

\bibitem{26}
M.~J.~C. Sousa and H.~M. Moreira, ``A survey on the software maintenance
  process,'' in \emph{Proceedings. International Conference on Software
  Maintenance (Cat. No. 98CB36272)}, Nov 1998, pp. 265--274.

\bibitem{56}
M.~G. Mehrabi, A.~G. Ulsoy, and Y.~Koren, ``Reconfigurable manufacturing
  systems: Key to future manufacturing,'' \emph{Journal of Intelligent
  Manufacturing}, vol.~11, no.~4, pp. 403--419, Aug 2000.

\bibitem{31}
S.~Srivastva and S.~Dhir, ``Debugging approaches on various software processing
  levels,'' in \emph{2017 International conference of Electronics,
  Communication and Aerospace Technology (ICECA)}, vol.~2, 2017, Conference
  Proceedings, pp. 302--306.

\bibitem{30}
S.~Soomro, M.~R. Belgaum, Z.~Alansari, and M.~H. Miraz, ``Fault localization
  models in debugging,'' in \emph{2017 International Conference on Infocom
  Technologies and Unmanned Systems (Trends and Future Directions) (ICTUS)},
  Dec 2017, pp. 57--62.

\bibitem{32}
\BIBentryALTinterwordspacing
M.~Perscheid, B.~Siegmund, M.~Taeumel, and R.~Hirschfeld, ``Studying the
  advancement in debugging practice of professional software developers,''
  \emph{Software Quality Journal}, vol.~25, no.~1, pp. 83--110, 2017. [Online].
  Available: \url{https://doi.org/10.1007/s11219-015-9294-2}
\BIBentrySTDinterwordspacing

\bibitem{33}
H.~Li, Y.~Xu, F.~Wu, and C.~Yin, ``Research of ``stub'' remote debugging
  technique,'' in \emph{2009 4th International Conference on Computer Science
  Education}, July 2009, pp. 990--994.

\bibitem{34}
\BIBentryALTinterwordspacing
M.~Lachwani and S.~J., ``Remote application debugging,'' October 2015, uS
  9,170,922 B1. [Online]. Available:
  \url{http://www.google.it/patents/US9170922B1}
\BIBentrySTDinterwordspacing

\bibitem{35}
C.~Wenyu, H.~Dongpu, T.~Dongcheng, and H.~Zongbo, ``A model of remote debugger
  supporting multiple types of connection,'' in \emph{2011 International
  Conference on Electronics, Communications and Control (ICECC)}, Sep. 2011,
  pp. 642--645.

\bibitem{36}
P.~J. Mosterman and J.~Zander, ``Cyber-physical systems challenges: a needs
  analysis for collaborating embedded software systems,'' \emph{Software {\&}
  Systems Modeling}, vol.~15, no.~1, pp. 5--16, Feb 2016.

\bibitem{37}
M.~Marra, E.~G. Boix, S.~Costiou, M.~Kerboeuf, A.~Plantec, G.~Polito, and
  S.~Ducasse, ``Debugging cyber-physical systems with pharo an experience
  report,'' in \emph{IWST 2017 - Proceedings of the 12th International Workshop
  on Smalltalk Technologies, in conjunction with the 25th International
  Smalltalk Joint Conference}, 2017, Conference Proceedings.

\bibitem{38}
\BIBentryALTinterwordspacing
Oracle, ``Java platformer debugger architecture,'' accessed: 2019-01-30.
  [Online]. Available:
  \url{https://docs.oracle.com/javase/8/docs/technotes/guides/jpda}
\BIBentrySTDinterwordspacing

\bibitem{39}
\BIBentryALTinterwordspacing
G.~Pothier, E.~Tanter, and J.~Piquer, ``Scalable omniscient debugging,'' in
  \emph{Proceedings of the 22Nd Annual ACM SIGPLAN Conference on
  Object-oriented Programming Systems and Applications}, ser. OOPSLA '07.\hskip
  1em plus 0.5em minus 0.4em\relax ACM, 2007, pp. 535--552. [Online].
  Available: \url{http://doi.acm.org/10.1145/1297027.1297067}
\BIBentrySTDinterwordspacing

\bibitem{40}
\BIBentryALTinterwordspacing
M.~Docs, ``Debug apps using visual studio - visual studio,'' accessed:
  2019-01-31. [Online]. Available:
  \url{https://docs.microsoft.com/en-us/visualstudio/debugger/debugger-feature-tour?view=vs-2017}
\BIBentrySTDinterwordspacing

\bibitem{41}
\BIBentryALTinterwordspacing
------, ``Remote debugging - visual studio,'' accessed: 2019-01-31. [Online].
  Available:
  \url{https://docs.microsoft.com/en-us/visualstudio/debugger/remote-debugging?view=vs-2017}
\BIBentrySTDinterwordspacing

\bibitem{42}
J.~Mickens, ``Rivet: Browser-agnostic remote debugging for web applications,''
  in \emph{Proceedings of the 2012 USENIX Conference on Annual Technical
  Conference}, ser. USENIX ATC'12.\hskip 1em plus 0.5em minus 0.4em\relax
  USENIX Association, 2012, pp. 30--30.

\bibitem{43}
A.~Sivieri, ``Eliot: A programming framework for the internet of things,''
  2014, phD thesis, politecnico di milano.

\bibitem{44}
A.~Sivieri, L.~Mottola, and G.~Cugola, ``Building internet of things software
  with eliot,'' \emph{Computer Communications}, vol. 89-90, pp. 141 -- 153,
  2016, internet of Things: Research challenges and Solutions.

\bibitem{45}
D.~Rathnayake, A.~Wickramarachchi, V.~Mallawaarachchi, D.~Meedeniya, and
  I.~Perera, ``A realtime monitoring platform for workflow subroutines,'' in
  \emph{2018 18th International Conference on Advances in ICT for Emerging
  Regions (ICTer)}, 2018, Conference Proceedings, pp. 41--47.

\bibitem{46}
D.~Gunter, E.~Deelman, T.~Samak, C.~Brooks, M.~Goode, G.~Juve, G.~Mehta,
  P.~Moraes, F.~Silva, M.~Swany, and K.~Vahi, ``Online workflow management and
  performance analysis with stampede,'' in \emph{2011 7th International
  Conference on Network and Service Management, CNSM 2011}, 01 2011, pp. 1--10.

\bibitem{47}
\BIBentryALTinterwordspacing
Node-Red, ``Node-red: About,'' accessed: 2019-02-5. [Online]. Available:
  \url{https://nodered.org/about/}
\BIBentrySTDinterwordspacing

\bibitem{48}
\BIBentryALTinterwordspacing
GDB, ``Debugging with gdb - the gdb remote serial protocol,'' accessed:
  2019-02-5. [Online]. Available:
  \url{https://cs.baylor.edu/~donahoo/tools/gdb/gdb.html\#SEC106}
\BIBentrySTDinterwordspacing

\bibitem{49}
\BIBentryALTinterwordspacing
Oracle, ``Java debug wire protocol,'' accessed: 2019-02-5. [Online]. Available:
  \url{https://docs.oracle.com/javase/8/docs/technotes/guides/jpda/jdwp-spec.html}
\BIBentrySTDinterwordspacing

\bibitem{dias17}
A.~Ramadas, G.~Domingues, J.~P. Dias, A.~Aguiar, and H.~S. Ferreira, ``Patterns
  for things that fail,'' in \emph{Proceedings of the 24th Conference on
  Pattern Languages of Programs}, ser. PLoP '17.\hskip 1em plus 0.5em minus
  0.4em\relax USA: The Hillside Group, 2017.

\bibitem{57}
W.~A.~Gruver and J.~Boudreaux, \emph{Intelligent Manufacturing:: Programming
  Environments for CIM}, 01 1993.

\bibitem{58}
\BIBentryALTinterwordspacing
C.~Manufacturing, ``Critical manufacturing - critical manufacturing mes |
  integrated manufacturing execution system,'' accessed: 2019-02-1. [Online].
  Available:
  \url{http://www.criticalmanufacturing.com/en/critical-manufacturing-mes/overview}
\BIBentrySTDinterwordspacing

\bibitem{59}
\BIBentryALTinterwordspacing
------, ``Critical manufacturing - connect iot,'' accessed: 2019-02-1.
  [Online]. Available:
  \url{http://www.criticalmanufacturing.com/en/critical-manufacturing-mes/connect-iot}
\BIBentrySTDinterwordspacing

\bibitem{29}
\BIBentryALTinterwordspacing
Tecon, ``Opc ua server,'' accessed: 2019-05-02. [Online]. Available:
  \url{https://www.tecon.cz/pdf/OPC\_UA\_UserManual\_en.pdf}
\BIBentrySTDinterwordspacing

\end{thebibliography}

\end{document}